\documentclass[letterpaper, 10 pt, conference]{ieeeconf}  

\IEEEoverridecommandlockouts                              

\usepackage{amsmath} 

\usepackage{amsthm}   
\usepackage{amssymb}  

\usepackage{mathtools}
\usepackage{pgf}
\usepackage[acronym]{glossaries} 
\usepackage{booktabs}
\usepackage{url}
\usepackage{hyperref}
\usepackage{cleveref}
\usepackage{graphicx}
\usepackage{subcaption}
\usepackage{multirow}

\Crefname{figure}{Fig.}{Figs.}

\newtheoremstyle{stdthm}
  {2pt}
  {2pt}
  {\itshape}
  {}
  {\bfseries\upshape}
  {.}
  {\,}
  {\thmname{#1}~\thmnumber{#2}
   \thmnote{\ \textnormal{[#3]}}}

\newtheorem{proposition}{Proposition}
\newtheorem{theorem}{Theorem}
\newtheorem{assumption}{Assumption}

\newacronym{MPC}{MPC}{Model Predictive Control}
\newacronym{EDS}{EDS}{Exponential Decay of Sensitivities}
\newacronym{NLP}{NLP}{Nonlinear Program}
\newacronym{LICQ}{LICQ}{Linear Independence Constraint Qualification}
\newacronym{SSOSC}{SSOSC}{Strong Second-Order Sufficient Conditions}
\newacronym{KKT}{KKT}{Karush-Kuhn-Tucker}
\newacronym{SQP}{SQP}{Sequential Quadratic Programming}
\newacronym{QP}{QP}{Quadratic Program}
\newacronym{DARE}{DARE}{Discrete Algebraic Riccati Equation}
\newacronym{LQR}{LQR}{Linear Quadratic Regulator}
\newacronym{ODE}{ODE}{Ordinary Differential Equation}
\newacronym{IRK}{IRK}{Implicit Runge-Kutta}
\newacronym{ERK}{ERK}{Explicit Runge-Kutta}
\newacronym{VDP}{VDP}{Van Der Pol}
\newacronym{MTS-MPC}{MTS-MPC}{Multi-Timescale MPC}

\title{\LARGE \bf
Multi-Timescale Model Predictive Control for Slow-Fast Systems
}

\author{Lukas Schroth$^{1,*}$, Daniel Morton$^{2, 3}$, Amon Lahr$^{1}$, Daniele Gammelli$^{2}$, \\ Andrea Carron$^{1}$ and Marco Pavone$^{2,4}$
\thanks{$^{1}$Institute for Dynamic Systems and Control, ETH Zürich}%
\thanks{$^{2}$Department of Aeronautics and Astronautics, Stanford University}%
\thanks{$^{3}$Department of Mechanical Engineering, Stanford University}%
\thanks{$^{4}$NVIDIA Research}%
 \thanks{$^{*}$Corresponding author: {\tt\small lschroth@ethz.ch}}%
\thanks{Daniel Morton was supported by a NASA Space Technology Graduate Research Opportunity}
}

\begin{document}

\maketitle

\thispagestyle{empty}
\pagestyle{empty}

\begin{abstract}

\gls{MPC} has established itself as the primary methodology for constrained control, enabling autonomy across diverse applications. While model fidelity is crucial in \gls{MPC}, solving the corresponding optimization problem in real time remains challenging when combining long horizons with high-fidelity models that capture both short-term dynamics and long-term behavior. Motivated by results on the \gls{EDS}, which imply that, under certain conditions, the influence of modeling inaccuracies decreases exponentially along the prediction horizon, this paper proposes a multi-timescale \gls{MPC} scheme for fast-sampled control. Tailored to systems with both fast and slow dynamics, the proposed approach improves computational efficiency by i) switching to a reduced model that captures only the slow, dominant dynamics and ii) exponentially increasing integration step sizes to progressively reduce model detail along the horizon. We evaluate the method on three practically motivated robotic control problems in simulation and observe speed-ups of up to an order of magnitude.\setcounter{footnote}{4}\footnote{Our code is available at \url{https://github.com/l-m-schroth/multi-timescale-mpc}}

\end{abstract}

\makeatletter
\global\@topnum\z@   
\global\@botnum\z@   
\makeatother
\begin{figure}[t]
  \centering
  \includegraphics[width=\linewidth]{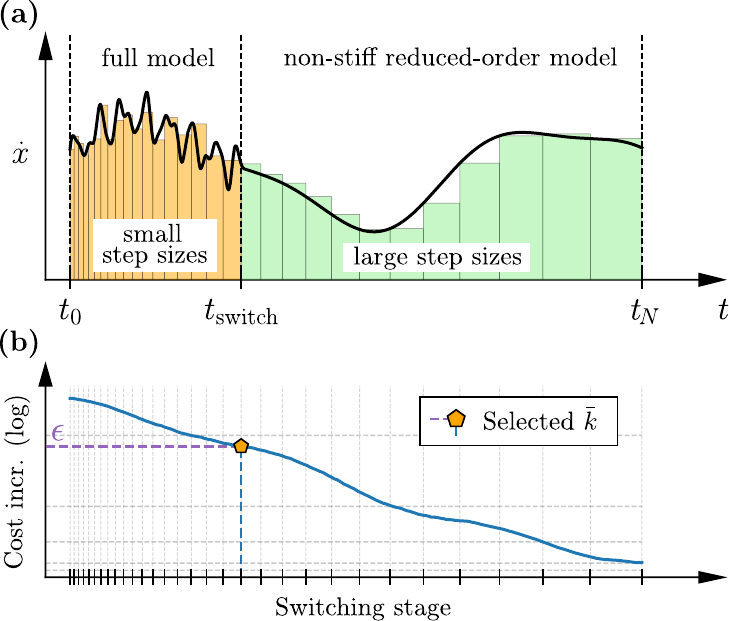} 
  \caption{Multi-Timescale MPC for slow-fast systems. 
    (a) The prediction horizon is split into two phases—first using the full model, then switching to a reduced model of the slow dynamics, enabling efficient discretization with exponentially increasing step sizes and explicit integration schemes. \\
    (b) Tuning of the switching stage: for a given step size profile, sweep over candidate switching stages and select the first stage for which the closed-loop cost increase relative to the full-resolution baseline is below a tolerance $\epsilon$.}
  \label{fig:cover_figure}
\end{figure}

\glsresetall 

\section{INTRODUCTION}

\gls{MPC}—an optimization-based control strategy that computes control inputs over a receding horizon—has been employed successfully for high-performance control of constrained systems across diverse applications, including process control~\cite{qin1997overview}, power electronics~\cite{geyer2016model}, wind turbine control~\cite{schlipf2013nonlinear}, and robotics~\cite{bangura2014real, bjelonic2022offline}. At each control step, only the first input of the optimized sequence is applied before the problem is re-solved, yielding a feedback mechanism that accounts for new measurements and disturbances.

For \gls{MPC} performance, it is crucial to plan sufficiently far ahead in order to consider the long-term effects of a control action. Further, the model must accurately reflect the system's dynamics to ensure reliable predictions. However, especially in robotics, increasing model fidelity often introduces fast dynamics with short time scales. For example, this occurs when modeling actuators, whose dynamics evolve much faster than those of the overall system, or when capturing high-frequency oscillations that arise in flexible joint manipulators~\cite{de2016robots}. The resulting systems contain multiple time scales; throughout this work, we refer to them as \emph{slow-fast systems}. Beyond robotics, time-scale separation also arises in other domains, including modern power systems with a high share of inverter-based resources~\cite{muntwiler2024stiffness}, wastewater treatment~\cite{brdys2008hierarchical}, and chemical process control~\cite{baldea2007control}.

Real-time feasibility of \gls{MPC} for slow-fast systems is often challenging. The slow dynamics necessitate long prediction horizons to accurately capture long-term behavior, whereas the stiffness associated with the fast modes requires implicit integration and small step sizes for accurate modeling. 
Altogether, this results in large-scale optimization problems that are computationally demanding, which is particularly restrictive in robotic applications where control must run at high frequency on resource-limited embedded hardware. 

Since only the first control input \(u_0\) is applied to the plant in \gls{MPC}, the closed-loop performance is primarily determined by \(u_0\), and modeling accuracy is most critical in those stages of the horizon that most strongly influence this initial input. Recently established results on the \gls{EDS} in optimal control show that, under mild regularity and second-order conditions, the influence of model perturbations on the optimal \(u_0\) decays exponentially with their distance along the prediction horizon \cite{na2020exponential, na2022convergence, shin2022exponential}. Motivated by this insight, we target the problem of reducing the computational burden of \gls{MPC} for slow-fast systems without compromising control fidelity and propose a practical approach that helps making \gls{MPC} with detailed multi-timescale models feasible in real time.

The contributions of this paper are as follows:
\begin{itemize}
    \item We introduce \gls{MTS-MPC}, a computationally efficient approach for controlling slow-fast systems. The method progressively increases the integration step size and transitions along the prediction horizon from a full, stiff model with implicit integration to a reduced-order, slow model with explicit integration. The proposed approach is visualized in \Cref{fig:cover_figure}.
    \item We evaluate \gls{MTS-MPC} in simulation against several \gls{MPC} formulations on three robotic systems: a differential drive robot, a drone, and a simplified robotic trunk, showing up to 16× speed-ups over a conventional full-resolution \gls{MPC} without compromising control performance. 
    \item We release a lightweight interface on top of \texttt{acados} that reduces the creation and evaluation of embedded \gls{MTS-MPC} solvers to a few lines of code.
\end{itemize}

\section{RELATED WORK}

\subsection{MPC with multiple models and non-uniform step sizes}

The idea of using multiple models within a single \gls{MPC} formulation has been explored in different domains, often demonstrating promising results in terms of computational efficiency and control performance. In \cite{li2021model}, a multi-phase scheme is proposed for efficient trajectory optimization of a 2D quadruped, employing full-order dynamics in the near term and a simplified, reduced-order model for longer-term predictions. The authors report near-identical performance compared to using the full model throughout, while cutting computation times in half. Building on this idea, highly dynamic quadruped locomotion is achieved in \cite{li2024cafe}, by formulating a multi-phase \gls{MPC} problem that tracks a reference trajectory generated offline.
An adaptive approach is proposed in \cite{norby2024adaptive}, where the model complexity is varied by identifying regions where the trajectory lies on a manifold, for which the complex model can be simplified to a lower-dimensional model. 
Beyond quadrupeds, multi-model \gls{MPC} has also been applied to humanoid systems \cite{li2024continuous, wang2021multi}. For instance, in \cite{li2024continuous}, models of different fidelity are used for different phases of a jumping maneuver with bipedal robots.

Applications of multi-model \gls{MPC} beyond legged robotics remain relatively rare, but some notable examples exist in process control \cite{kameswaran2012multi} and autonomous racing, where multi-model \gls{MPC} enables longer horizons without compromising real-time feasibility and results in faster lap times on hardware \cite{laurense2021long}. Similarly, non-uniform time discretizations that use short steps in the near term and longer steps farther ahead have been shown to reduce computational demand without degrading control performance~\cite{tan2016model} and, at fixed computational complexity, to improve path-following and obstacle-avoidance of autonomous vehicles~\cite{kim2021model}.

Recursive feasibility guarantees under a single model switch have been addressed using tube-based constraint tightening \cite{bathge2016exploiting}. The approach in \cite{bathge2016exploiting} has been generalized in \cite{brudigam2021model} to account for model switches where the integration step sizes of the individual models differ, while remaining constant within each model phase.

\subsection{Existing approaches for slow-fast systems}

While the literature clearly highlights the need for control strategies capable of handling multiple time scales \cite{brdys2008hierarchical, baldea2007control, van2009time, muntwiler2024stiffness}, existing approaches often face limitations.

The most prominent family of approaches to control slow-fast systems consists of hierarchical decomposition schemes, which address different time scales via separate control loops \cite{scattolini2009architectures, picasso2010mpc, brdys2008hierarchical, van2009time, baldea2007control}. Typically, high-level controllers, running at a low frequency, use low-complexity formulations based on reduced-order models to compute reference control inputs that achieve a desirable long-term behavior of the plant \cite{scattolini2009architectures}. These inputs serve as references for a lower-level tracking controller, which runs at a higher rate and considers the full model \cite{scattolini2009architectures}. 

As evidenced by \cite{brdys2008hierarchical, baldea2007control, van2009time}, hierarchical approaches are often tailored to specific applications rather than being general-purpose solutions \cite{scattolini2009architectures}. Further, hierarchical schemes do not solve the original optimization problem directly due to the decomposition. Entirely neglecting the fast dynamics in the outer-loop optimization may lead to suboptimal performance or safety risks in tasks where these dynamics are important—a limitation shared by approaches that rely on \gls{MPC} using reduced-order models of the slow dynamics over the entire horizon \cite{alora2023data}. Due to the hierarchical decomposition, it is also unclear how to achieve a desired balance between competing objectives at different time scales.

A different class of approaches aimed at handling long-horizon and large-scale problems in \gls{MPC} focuses on decomposition at the optimization level \cite{shin2019parallel, giselsson2013accelerated}. Commonly, the time domain is decomposed into several subdomains with partially overlapping regions, allowing for parallel computation of local subproblems, whose solutions are then connected to construct globally consistent primal-dual trajectories \cite{shin2019parallel}. However, distributed approaches to large-scale convex optimization have been found to be non-competitive compared to solving the problem centrally using structure-exploiting solvers \cite{kozma2015benchmarking} and they are often not readily available in open-source model predictive control software.

Closest to our approach are time coarsening strategies, which aim to reduce the number of optimization variables in the underlying optimal control problem. In \cite{shin2021diffusing}, the authors propose reducing problem size using a stage aggregation approach tailored to linear programming-based predictive control problems — an approach that is, by design, limited to this specific problem class. Another strategy is to fix the inputs or their derivatives to be constant over several time steps (also referred to as move-blocking) \cite{cagienard2007move}, which reduces only the number of optimization variables corresponding to the inputs, without addressing those related to the states. Another work speeds up \gls{MPC} computation for multi-timescale processes by choosing small time intervals in the near future and large intervals for distant predictions \cite{tan2016model}. However, large time steps for dynamically stiff problems require expensive implicit integration and are prone to numerical errors.

In summary, existing methods either (i) consider fast dynamics only in the inner-loop tracking while neglecting them in the outer-loop planning problem (hierarchical schemes); (ii) reduce only decision variables associated with inputs (move blocking), leaving state variables unaddressed; or (iii) can suffer from numerical errors caused by combining stiff dynamics with large integration steps. In response to these challenges, we present an MPC approach for slow-fast systems that maintains full-dynamics awareness without hierarchical decomposition and reduces problem size for efficient centralized solution, while avoiding numerical issues associated with large integration steps.

\section{METHODOLOGY}
This section develops the proposed \gls{MTS-MPC} method and is structured as follows: we first outline the standard MPC formulation used as a baseline, then present the proposed approach to reduce the associated computational burden, and finally provide a practical strategy for tuning.

\subsection{Baseline Formulation}
We consider systems governed by
\begin{equation}
\dot{x}(t) = f_{\text{full}}(x(t), u(t)),
\end{equation}
and formulate the discrete-time MPC problem using direct multiple shooting \cite{bock1984multiple}.
The horizon $T$ is partitioned into $N$ intervals, with both the state values at the shooting nodes and the piecewise constant input values treated as optimization variables.
The dynamics are integrated forward within each interval, resulting in equality constraints that ensure consistency between the discrete decision variables and the underlying continuous dynamics. Slow-fast systems are often numerically stiff, requiring implicit integration and small step sizes for stability and accuracy. Small integration steps, however, increase the number of shooting intervals $N$ needed for a fixed prediction horizon $T$, leading to a large and computationally demanding baseline formulation:
\begin{subequations}
\begin{align}
\label{eq:baseline_problem}
    \min_{\substack{\boldsymbol{x}, \boldsymbol{u}}} \quad &
        \ell_N(x_N) + \sum_{k=0}^{N-1} \ell(x_k, u_k) \\
    \text{s.t.} \quad &
        x_{k+1} = f_{\text{full}}^{\text{IRK}}(x_k, u_k), \quad k = 0,\dots,N-1, \\
    & c_k(x_k, u_k) \le 0, \quad k = 0,\dots,N-1, \\
    & c_N(x_N) \le 0, \\
    & x_0 = x(t).
\end{align}
\end{subequations}
where $f_{\text{full}}^{\text{IRK}}$ denotes the discrete-time model obtained using an implicit Runge-Kutta scheme. The state and input trajectories are represented by $\boldsymbol{x} = \{x_k\}_{k=0}^{N}$ and $\boldsymbol{u} = \{u_k\}_{k=0}^{N-1}$. The functions $\ell$, $\ell_N$, $c_k$, and $c_N$ define the stage and terminal cost and constraint terms, respectively, and $x(t)$ denotes the current system state.

\subsection{Proposed Method}

Many slow-fast systems exhibit clearly separable slow and fast dynamics. As a result, model decomposition into systems of different fidelity often arises naturally~\cite{scattolini2009architectures}. Even when this separation is not explicit, model-order reduction techniques have been proposed to alleviate stiffness and extract dominant slow dynamics~\cite{muntwiler2024stiffness, alora2023data}. We refer to the reduced model of the slow dynamics as
\begin{equation}
    \dot{x}_{\text{slow}}(t) = f_{\text{slow}}(x_{\text{slow}}(t), u_{\text{slow}}(t)).
\end{equation}

Motivated by this natural decomposition and the \gls{EDS}, our method accelerates \gls{MPC} for slow-fast systems by combining an exponentially increasing integration step size along the prediction horizon with a multi-model strategy, reflecting the intuition that fast dynamics are critical for short-term decisions, whereas slow dynamics dominate long-term behavior. We use the full stiff model in the early stages (up to switching stage $\bar{k}$, $0 \le \bar{k} \le N-1$) and switch to a reduced model of the slow dynamics integrated with an explicit scheme, resulting in the following discrete-time optimal control problem:
\begin{subequations}
\begin{align}
    \min_{\substack{\boldsymbol{x}, \boldsymbol{u}}} \quad &
        \ell_N(x_N)
        + \sum_{k=0}^{\bar{k}-1} \ell(x_k, u_k)
        + \sum_{k=\bar{k}}^{N-1} \ell_{\text{slow}}(x_k, u_k) \\
    \text{s.t.} \quad &
        x_{k+1} =
        \begin{cases}
            f_{\text{full}}^{\text{IRK}}(x_k, u_k, \Delta t_k), & k < \bar{k}, \\[3pt]
            f_{\text{slow}}^{\text{ERK}}(x_k, u_k, \Delta t_k), & k \geq \bar{k},
        \end{cases} \\[3pt]
    & c_k(x_k, u_k) \le 0, \quad k = 0,\dots,N-1, \\[3pt]
    & c_N(x_N) \le 0, \\[3pt]
    & x_{\bar{k}} = \phi(x_{\bar{k}}^-), \quad x_0 = x(t).
\end{align}
\end{subequations}
Here, the time steps are predefined by an exponential schedule $\Delta t_k = \Delta t_0\, \alpha^{k}$, $\alpha > 1$, and $\phi(\cdot)$ is a differentiable mapping between models that projects the state at the switching stage, $x_{\bar{k}}^-= f^{\text{IRK}}_{\text{full}}(x_{\bar{k} - 1}, u_{\bar{k} - 1})$, from the full-model state space onto the reduced-model state space. For notational simplicity, the same symbols $x_k$ and $u_k$ are used across both phases (despite potentially varying dimension). As demonstrated in \Cref{sec:experiments}, the proposed strategy achieves computational savings of up to an order of magnitude while preserving the performance of the baseline formulation.
More generally, our approach has the following advantages:
\begin{enumerate}
    \item \textbf{Significant reduction in problem size:} Achieved through both the exponential increase in integration step sizes and switching to a lower-dimensional model.
    \item \textbf{Avoidance of integration errors:} The use of an approximate, non-stiff model allows for accurate numerical integration with larger step sizes and cheaper integration schemes, often enabling a switch from implicit methods to explicit ones.
    \item \textbf{Simplified design and tuning effort:} Unlike hierarchical schemes, our approach directly approximates the original optimal control problem. Competing goals for slow and fast dynamics can simply be balanced within the objective function.
    \item \textbf{Ease of implementation:} Our method is compatible with existing open-source \gls{MPC} tools such as \texttt{acados} \cite{verschueren2022acados, frey2024multi}, allowing for straightforward problem specification and efficient solution using state-of-the-art structure-exploiting solvers.
\end{enumerate}

While the \gls{EDS} describes a continuous decrease in the influence of perturbations along the prediction horizon, thereby suggesting a gradual reduction of model order beyond time-scale separation, we found that increasing the integration step size offers a more practical and straightforward way to coarsen the optimization problem. Our motivation to consider exponentially increasing step sizes is twofold. First, it aligns naturally with the exponential rate of the sensitivity decay. Second, in the context of linear optimal control problems, it was shown in \cite{shin2021diffusing} that aggregating shooting nodes at an exponential rate yields the best trade-off between closed-loop cost and computational effort compared to alternatives.

\subsection{Tuning}
\begin{figure}[t]
  \centering
  \includegraphics[width=1.0\linewidth]{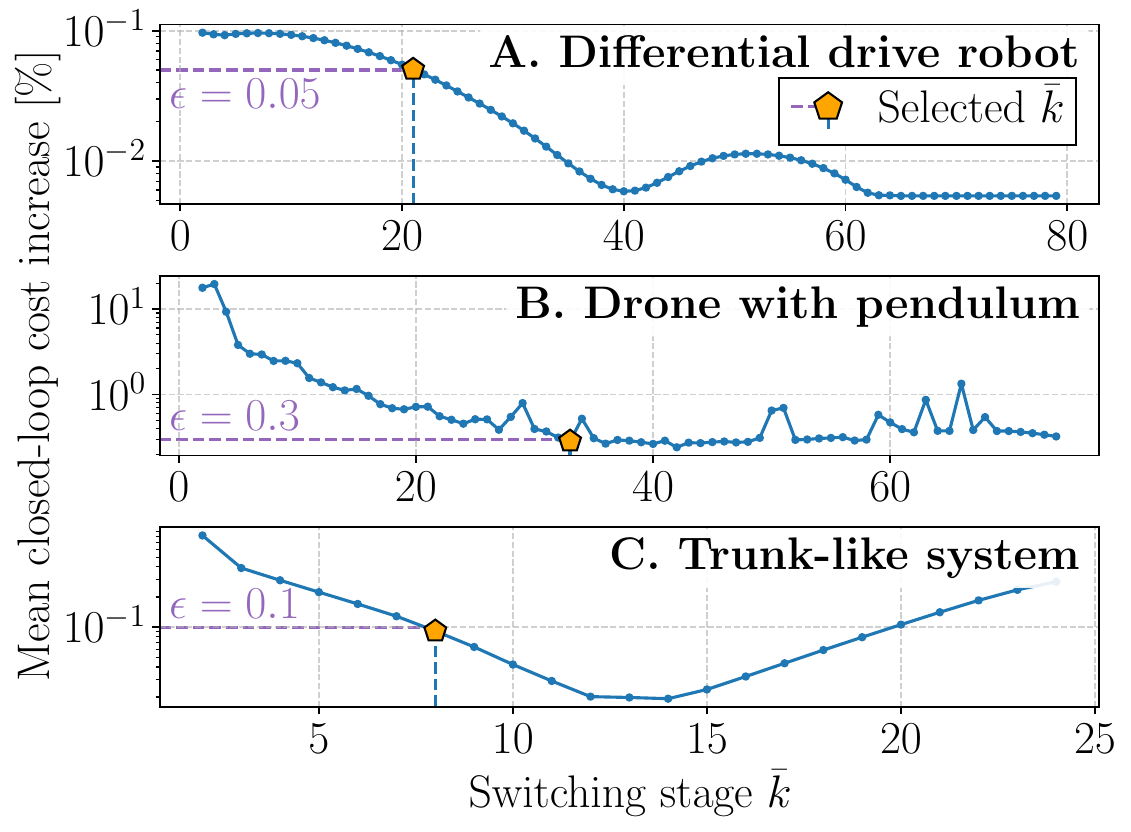} 
  \caption{Mean closed-loop cost increase (relative to the full-resolution baseline) versus switching stage $\bar{k}$. The orange pentagon indicates the selected switching stage.}
  \label{fig:switching_stage_selection}
\end{figure}

The switching stage and step size schedule should be chosen based on the desired trade-off between accuracy and computational efficiency. The step size needs to increase sufficiently to reduce the number of shooting nodes. However, since constraints are enforced only at those nodes, overly coarse integration can cause violations, for instance by letting trajectories slip through obstacles. In our experiments, we use a tuning scheme that minimizes computational effort while preserving the closed-loop performance of the baseline.
We first fix a moderate step size schedule and then determine the switching stage by sweeping it from early to late in the horizon. Each configuration is evaluated in closed loop starting from the same initial state used in the experiments (a process which can be extended to multiple initial conditions). The first stage is selected for which the closed-loop cost increase becomes negligible, falling below a specified tolerance~$\epsilon$ relative to the baseline. If the desired accuracy cannot be achieved for the given step size schedule, smaller step sizes are chosen and the sweep is repeated. Figure~\ref{fig:switching_stage_selection} illustrates this process, showing the closed-loop cost increase versus switching stage for all examples, along with the selected thresholds~$\epsilon$ and resulting switching stages. Interestingly, for the trunk system (and to a much lesser extent for the drone), the closed-loop cost increases again when switching in the final stages. We attribute this behavior to integration errors that occur when the stiff full model is simulated with large step sizes, whereas switching earlier to the non-stiff approximate model mitigates these errors and can even improve closed-loop performance. 

\section{EXPERIMENTAL RESULTS}
\label{sec:experiments}
In this section, we evaluate \gls{MTS-MPC} on several examples of slow-fast systems. To demonstrate its effectiveness, we compare the following \gls{MPC} formulations:

\begin{itemize}
    \item[0)] \textbf{Baseline:} Using the full model with constant small step sizes and a long prediction horizon  
    (All other \gls{MPC} formulations are approximations of this baseline; apart from the explicitly listed modifications below, all parameters are kept identical)
    \item[1)] \textbf{Myopic \gls{MPC}:} Using a shorter prediction horizon, resulting in fewer shooting nodes
    \item[2)] \textbf{Lower frequency:} Using larger step sizes, reducing controller frequency accordingly
    \item[3)] \textbf{Approximate model only:} Using only the reduced-order model (only considered in examples where full and approximate models share the same inputs)
    \item[4)] \textbf{Full model with increasing step sizes:} Reducing the number of shooting intervals via exponential step size growth (same step size schedule as for 6))
    \item[5)] \textbf{Model switching only:} Switching from the full to the reduced-order model without increasing step sizes
    \item[6)] \textbf{\gls{MTS-MPC}:} Combining model switching with exponentially increasing step sizes
\end{itemize}

We implement all predictive controllers using the open-source \gls{SQP} framework \texttt{acados} \cite{verschueren2022acados}. Multi-model formulations are implemented via its multi-phase interface \cite{frey2024multi}. The quadratic programs arising in \gls{SQP} are solved using the state-of-the-art solver \texttt{HPIPM} \cite{frison2020hpipm}. Closed-loop simulations are performed by integrating the system \glspl{ODE} using small step sizes and implicit integration. All controllers are applied with frequency $1/\Delta t_0$, where $\Delta t_0$ is the initial step size. Depending on the example-specific need for implicit integration, either implicit or explicit integration is used for discretizing the continuous-time optimal control problem via multiple shooting. The standard $\texttt{acados}$ \gls{IRK} and \gls{ERK} integrators used implement an eighth-order Gauss-Legendre method and a fourth-order classical Runge-Kutta method, respectively. We emphasize that all controllers share the same prediction horizon $T$ as the baseline, with the natural exception of the myopic controller. Complete controller configurations and physical parameters for all tasks are summarized in the appendix. We consider the following control tasks.

\begin{figure*}[t]          
  \centering
  \includegraphics[width=\textwidth]{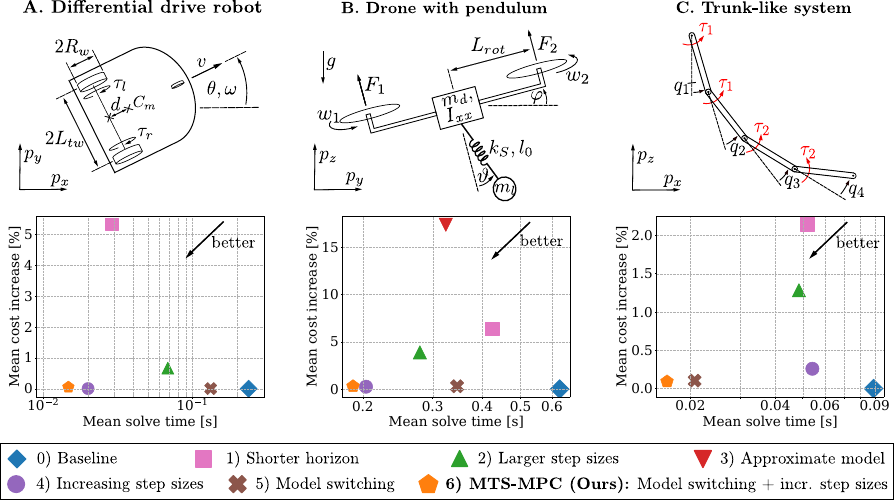}%
  \caption{System schematics and Pareto frontiers. MTS-MPC (orange) consistently outperforms all baselines in the trade-off between mean closed-loop stage costs and computational efficiency across three slow-fast robotic systems.}
  \label{fig:main}
  \vspace{-3mm} 
\end{figure*}

\subsection{Differential drive robot}
\label{sec:Diff-drive}

We consider a differential-drive robot tasked with reaching a desired goal state. The full model includes the actuator dynamics of the DC motors. Here, the electrical subsystem evolves significantly faster than the robot’s mechanical state, introducing stiffness. The reduced model omits these fast actuator states and is directly actuated via wheel torques.
The optimal control problem closely follows \cite{frey2024multi}, with a modified objective. The differential-drive model, with and without actuators, originates from \cite{dhaouadi2013dynamic}.

\subsubsection{Simplified model}

The reduced-order model omits actuator dynamics and is shown in \Cref{fig:main}-A. The state is 
\begin{equation}
x_{\text{slow}} = [p_x,\, p_y,\, v,\, \theta,\, \omega]^\top,
\end{equation}
with \( p_x, p_y \in \mathbb{R} \) representing the robot's position in the world frame, \( v \) the linear velocity, \( \theta \) the heading, and \( \omega \) the angular velocity. The control inputs are given by torques applied to the left and right wheel:
\begin{equation}
u_{\text{slow}} = [\tau_r,\, \tau_l]^\top.
\end{equation}
The system dynamics are described by the following \gls{ODE}:
\begin{equation}
\dot{x}_{\text{slow}} =
\begin{bmatrix}
v \cos \theta \\
v \sin \theta \\
\frac{a_1 + m_c d \omega^2}{m_{tot} + a_2} \\
\omega \\
\frac{L_{tw} a_3 - m_c d \omega v}{I_{zz} + L_{tw}^2 a_2}
\end{bmatrix},
\end{equation}
with the auxiliary variables defined as:
\begin{equation}
a_1 = \frac{\tau_r + \tau_l}{R_{w}}, \quad
a_2 = \frac{2 I_w}{R_{w}^2}, \quad
a_3 = \frac{\tau_r - \tau_l}{R_{w}}.
\end{equation}

\subsubsection{Full model}
To capture actuator behavior, we augment the state with motor currents:
\begin{equation}
x = [p_x,\, p_y,\, v,\, \theta,\, \omega,\, I_r,\, I_l]^\top,
\end{equation}
where \( I_r \) and \( I_l \) are the currents in the motors driving the right and left wheels, respectively. The control inputs are now the motor voltages:
\begin{equation}
u = [V_r,\, V_l]^\top.
\end{equation}
The mechanical dynamics match those of the simplified model, but torques are expressed in terms of motor currents:
\begin{equation}
\tau_r = K_1 I_r, \quad \tau_l = K_2 I_l.
\end{equation}
The electrical subsystem evolves according to:
\begin{equation}
\begin{bmatrix}
\dot{I}_r \\
\dot{I}_l
\end{bmatrix}
=
\begin{bmatrix}
-\frac{K_1 \psi_1 - R_{\mathrm{act}} I_r + V_r}{L_{\mathrm{act}}} \\
-\frac{K_2 \psi_2 - R_{\mathrm{act}} I_l + V_l}{L_{\mathrm{act}}}
\end{bmatrix},
\end{equation}
with:
\begin{align}
\psi_1 &= \frac{\dot{p}_x \cos \theta + \dot{p}_y \sin \theta + L_{tw} \omega}{R_{w}}, \\
\psi_2 &= \frac{\dot{p}_x \cos \theta + \dot{p}_y \sin \theta - L_{tw} \omega}{R_{w}}.
\end{align}

\subsubsection{Optimal control problem}

For the full model, the stage cost incentivizes tracking of the zero state via a quadratic penalty:
\begin{equation}
\ell(x, u) = x^\top Q
x.
\end{equation}
For parts of the horizon where the simplified model is considered (e.g., in the second phase of the proposed approach), we additionally penalize the control inputs to account for the torques in the objective:
\begin{equation}
\ell_{\text{slow}}(x, u) = x^\top \hat{Q} x + u^\top \hat{R} u, 
\end{equation}
while the terminal cost uses the same quadratic penalty on the state only.
State constraints for both models are:
\begin{equation}
0 \leq v \leq 1, \quad -0.5 \leq \omega \leq 0.5,
\end{equation}
while control input bounds depend on the model variant. For the actuator-based model we enforce
\begin{equation}
-10 \leq V_r, V_l \leq 10,
\end{equation}
Similarly for the simplified model, it must hold that
\begin{equation}
-60 \leq \tau_r, \tau_l \leq 60.
\end{equation}

\subsection{Oscillation-aware drone control}
\label{sec:oscillations_aware_drone_control}

The second task considers a 2D drone with a fragile load suspended by a stiff spring, modeling material-induced oscillations. Comparable scenarios include a camera or sensor mounted on the drone that must remain stable during flight. The goal is to reach a target position while minimizing load acceleration, requiring active damping of pendulum oscillations. The full model includes spring extension as a state, while the approximate model replaces it with a rigid pendulum of fixed length.

\subsubsection{Full model}

The setup is visualized in \Cref{fig:main}-B.
The drone moves in the \((y,z)\)-plane and carries a mass~\(m_l\) suspended by a spring with rest length~\(l_0\) and stiffness~\(k_S\). The drone itself has a mass of $m_d$ and a moment of inertia of $I_{xx}$. The full model includes the time-varying spring length \(r(t)\), with state and control defined as:
\begin{align}
x &= [\, p_y\;\; p_z\;\; \varphi\;\; r\;\; \vartheta\;\; \dot{p}_y\;\; \dot{p}_z\;\; \dot{\varphi}\;\; \dot{r}\;\; \dot{\vartheta}\;\; w_1\;\; w_2\,]^\top, \\
u &= [\, \dot{w}_1\;\; \dot{w}_2\,]^\top.
\end{align}
where \(p_y, p_z\) denote the drone’s position, \(\varphi\) its roll, \(r\) the spring length, \(\vartheta\) the pendulum angle, and \(w_1, w_2\) the rotor speeds. For simplicity, we assume a linear thrust model $w_i$: \(F_i = \kappa\, w_i\), which permits bidirectional force generation as found in drones capable of inverted or agile flight \cite{jothiraj2019enabling}. The equations of motion, derived using Lagrangian dynamics, are provided in the appendix.
\subsubsection{Simplified model} 
The simplified model assumes a rigid pendulum of fixed length $l_0$, effectively removing the spring extension $r$ and its rate $\dot{r}$ from the reduced state $x_{\text{slow}}$.
For this example, we deviate from the integrator switching used in our method: since no notable performance difference was observed between \gls{ERK} and \gls{IRK} when integrating the full model, all controllers are implemented using \gls{ERK} to ensure a fair comparison of computation times.

\subsubsection{Optimal control problem}
The cost is identical for both models, penalizing deviation from the goal, load acceleration, and control effort. Defining
\begin{equation}
    \Delta x_{\text{pos}} :=  
    \begin{bmatrix} p_y \\ p_z \end{bmatrix} - 
    \begin{bmatrix} p_{y,\text{goal}} \\ p_{z,\text{goal}} \end{bmatrix}, \quad
    a_{\text{load}} :=  
    \begin{bmatrix} \ddot{p}_{y,\text{load}} \\ \ddot{p}_{z,\text{load}} \end{bmatrix},
\end{equation}
the stage costs $\ell(x,u)$ and $\ell_{\text{slow}}(x,u)$ are both given by
\begin{equation}
\label{eq:stage_costs}
    \Delta x_{\text{pos}}^\top Q \, \Delta x_{\text{pos}} +
    a_{\text{load}}^\top Q_{\text{load}} \, a_{\text{load}} +
    u^\top R \, u.
\end{equation}
We further enforce box constraints on the motor spin rates $w_1$, $w_2$, the control inputs $u_1$, $u_2$, the admissible drone pitch angle $\varphi$, and the thrusts $F_1$, $F_2$:
\begin{align}
-10 &\leq w_1,\, w_2 \leq 10, \quad & -5  &\leq u_1,\, u_2 \leq 5, \nonumber \\
-0.7 &\leq \varphi \leq 0.7, \quad & -15 &\leq F_1,\, F_2 \leq 15.
\end{align}

\begin{figure*}[t]          
  \centering
  \includegraphics[width=\textwidth]{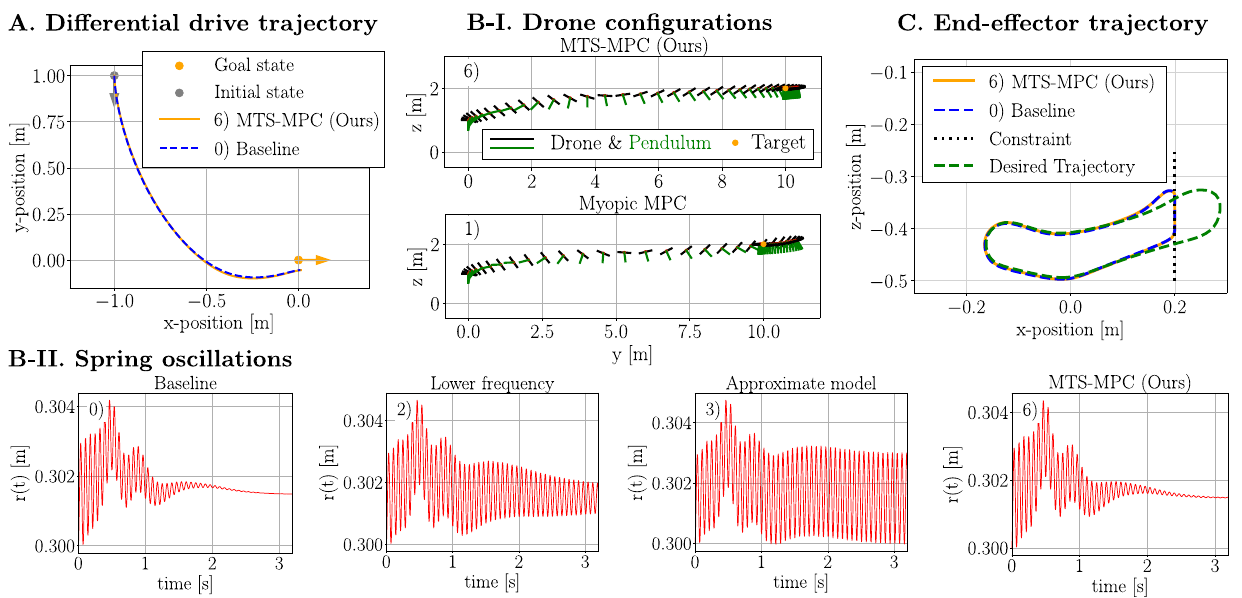}%
  \caption{Closed-loop trajectories: MTS-MPC achieves the same performance as the full-resolution baseline while being 16×, 3.3×, and 5.4× more computationally efficient for the differential-drive, drone, and trunk-like systems, respectively.}
  \label{fig:closed_loop}
  \vspace{-3mm} 
\end{figure*}

\subsection{Trajectory tracking with a simplified robotic trunk}
This example studies a 2D system approximating the dynamics of a flexible robotic trunk. Such systems are known to comprise both slow and fast dynamic components \cite{buurmeijer2025taming}, leading to stiffness. The full model consists of rigid links connected by hinge joints with stiffness and damping; the reduced model treats the end-effector as a point mass with second-order dynamics. Although the experiments use a 16-link system, a four-link version is shown in \Cref{fig:main}-C for clarity. The task is to track a desired end-effector trajectory.

\subsubsection{Full model}
The state is defined in terms of generalized coordinates and consists of joint angles and their corresponding angular velocities $ x = [q\;\; \dot{q}]^\top $.
 Actuation is applied as joint torques, with identical torques enforced across the upper and lower halves of the system. This setup reflects the actuation by two pairs of antagonistic tendons: one pair terminating at the midpoint of the chain, while the other extends to the last link. Since (idealized) tendon contraction generates equal tension at all attachment points along the link, the resulting torques around the corresponding hinge joints remain consistent across each section.

\subsubsection{Simplified model}
The reduced model approximates the end-effector as a point mass with second-order dynamics:
\begin{equation}
x_{\text{slow}} = 
\begin{bmatrix} 
p^{\mathrm{ee}} \\ 
v^{\mathrm{ee}} 
\end{bmatrix},
\quad
\dot{x}_{\text{slow}} =
\begin{bmatrix} 
v^{\mathrm{ee}} \\ 
u_{\text{slow}} 
\end{bmatrix},
\end{equation}
where $p^{\mathrm{ee}}$ is the position of the end-effector, $v^{\mathrm{ee}}$ its velocity, and $u_{\text{slow}}$ the control input corresponding to its acceleration.
The mapping $\phi$ from the full model to the point-mass model is straightforward: while the position of the end-effector can simply be computed based on the system's kinematics, the velocity \( v^{ee} \) is obtained via the position Jacobian computed through automatic differentiation.


\subsubsection{Optimal control problem}
Let \( \Psi(t) \in \mathbb{R}^2 \) denote the desired time-varying reference for the end-effector position. The stage cost penalizes deviation from this reference as well as control effort and is given by:
\begin{equation}
\ell(x, u) \coloneqq (p^\mathrm{ee} - \Psi(t_{\text{stage}}))^\top Q (p^\mathrm{ee} - \Psi(t_{\text{stage}})) + u^\top R u,
\end{equation}
Here $t_{\text{stage}}$ is the global time associated with each stage of the horizon, depending on the current time and the chosen step size schedule. For the point-mass model, $\ell_{\text{slow}}$ shares the same structure as $\ell$ but differs in the weighting of the control effort through a separately tuned $R$.

\subsection{Results}

\Cref{fig:main} visualizes the Pareto frontiers of closed-loop cost versus computation time for all examples. Across all three cases, the proposed approach consistently dominates the Pareto frontier, achieving the lowest computational cost among all methods tested while maintaining performance comparable to the baseline. In the differential-drive example, it achieves a speed-up of approximately 16x in total solve time and around 19x in time per \gls{SQP} iteration compared to the baseline, while incurring only a numerically negligible increase in closed-loop cost of \(5.0 \cdot 10^{-2}\%\). In the drone control task, it lowers solver time and per-iteration time to approximately 30\,\% and 27\,\% of the baseline, respectively, with a cost increase of just 0.29\,\%. In the robotic trunk example, computation time is reduced to 18.5\,\% of the baseline, while the closed-loop cost increases by only 0.091\,\%. The similarity in performance can also be observed in \Cref{fig:closed_loop}, where the closed-loop trajectories of the proposed approach closely match those of the baseline MPC.

In the differential-drive and drone examples, a comparison with the individual components of the proposed approach—4) using only an exponential increase in integration step size and 5) using only model switching—shows that most computational savings stem from the step size increase. In these cases, the fast dynamics are rather low-dimensional, so neglecting them yields only minor additional gains. In the robotic trunk example, however, model switching dominates. Since accurate trajectory tracking in this case does not require a long horizon, the benefit from step size increase is limited, while the dimensionality reduction from switching the full 32-dimensional model to a low-dimensional point-mass model leads to significant speed-ups. Further, simply increasing the integration step size for the full model leads to noticeable performance loss, and more aggressive step size schedules can even result in complete tracking failure. Switching to the reduced non-stiff model allows for larger integration steps without sacrificing accuracy.

In all examples, alternative approximation strategies (myopic horizon, coarse step sizes with the full stiff model, and using only the approximate model) result in clear performance degradation. In the drone control task (\Cref{fig:closed_loop}-B), increasing the step size and thereby lowering the control frequency 2), and using only the approximate model 3), both fail to sufficiently damp the pendulum oscillations (\Cref{fig:closed_loop}-B-II). The myopic \gls{MPC} 1), in contrast, manages to suppress oscillations but overshoots the goal state (\Cref{fig:closed_loop}-B-I). The proposed approach completes both subtasks successfully.

\section{CONCLUSIONS \& DISCUSSION}
Motivated by the recently established concept of \gls{EDS}, we propose \gls{MTS-MPC}, a novel method to address the computational challenges of predictive control in systems with both fast and slow dynamics. The approach combines an initial use of the full model with a switch to a reduced-order model that captures only the slow dynamics, enabling accurate integration with larger step sizes and computationally cheaper explicit schemes. Across several control tasks, the method achieves speed-ups of up to an order of magnitude while maintaining the performance of the full-model baseline. These results suggest that \gls{MTS-MPC} can help to enable real-time control of slow-fast systems and may make deployment on less expensive hardware possible. 

Future work will include applications to oscillation-aware \gls{MPC} for flexible joint manipulators and control of power systems with a large share of inverter-based resources, as well as hardware experiments to further validate the method. 
An open challenge is the development of schemes with recursive feasibility and stability guarantees when combining model switching with step size increase, especially for nonlinear \gls{MPC}. 



\bibliographystyle{IEEEtran}
\bibliography{bibliography}

\clearpage           

\section*{APPENDIX}

\subsection{Exponential Decay of Sensitivities}

\subsubsection{Simple example and visualization}
\label{sec:EDS_simple_example}

To build intuition for \gls{EDS}, we consider a simple setting with quadratic costs, linear dynamics affected by additive disturbances $p_k$, and no inequality constraints:

\begin{subequations}\label{eq:QP_EDS_example}
\begin{alignat}{2}
    \min_{x_k, u_k} \quad & x_N^T S x_N + \sum_{k=0}^{N-1} \left( x_k^T Q x_k + u_k^T R u_k \right)  \\
    \text{s.t.} \quad & x_{k+1} = A x_k + B u_k + p_k, \quad k \in [N-1] \\
    & x_0 = x(k),
\end{alignat}
\end{subequations}
where we use the compact notation $[N\!-\!1] := \{0, \ldots, N\!-\!1\}$. In this case, the sensitivities of the optimal control input to additive errors at different stages of the horizon can be derived in closed form.  

\begin{proposition}[Sensitivity of the Optimal Control Input]
\label{prop:sensitivities_motivating_example}
    Assume \( Q \succeq 0 \) and \( R \succ 0 \), and define the lifted system matrices:
    \begin{equation}
\begin{aligned}
    \mathcal{X} &= \begin{pmatrix}
        A \\
        A^2 \\
        \vdots \\
        A^N
    \end{pmatrix}, 
    \quad
    \mathcal{P} = \begin{pmatrix}
        I & 0 & 0 & \cdots & 0 \\
        A & I & 0 & \cdots & 0 \\
        A^2 & A & I & \cdots & 0 \\
        \vdots & \vdots & \vdots & \ddots & \vdots \\
        A^{N-1} & A^{N-2} & A^{N-3} & \cdots & I
    \end{pmatrix}, \\
    \mathcal{U} &= \begin{pmatrix}
        B & 0 & 0 & \cdots & 0 \\
        A\,B & B & 0 & \cdots & 0 \\
        A^2\,B & A\,B & B & \cdots & 0 \\
        \vdots & \vdots & \vdots & \ddots & \vdots \\
        A^{N-1}\,B & A^{N-2}\,B & A^{N-3}\,B & \cdots & B
    \end{pmatrix}.
\end{aligned}
\end{equation}
Furthermore, define the vectorized control input and disturbance as
    \begin{equation}
        U = \begin{pmatrix} u_0 \\ u_1 \\ \vdots \\ u_{N-1} \end{pmatrix}, \quad 
        P = \begin{pmatrix} p_0 \\ p_1 \\ \vdots \\ p_{N-1} \end{pmatrix}, 
    \end{equation}
    and the block-diagonal cost matrices as
    \begin{equation}
        \mathcal{Q} = \text{diag}(Q, Q, \dots, S), \quad \mathcal{R} = \text{diag}(R, R, \dots, R).
    \end{equation} 
    Then the optimal control inputs of the \gls{QP} \cref{eq:QP_EDS_example} satisfy
    \begin{equation}
        U^\star = -\bigl(\mathcal{U}^T \mathcal{Q} \mathcal{U} + \mathcal{R}\bigr)^{-1} 
        \mathcal{U}^T \mathcal{Q} \bigl(\mathcal{X}\,x_0 + \mathcal{P}\,P\bigr),
    \end{equation}
    and its sensitivity with respect to the disturbance vector \(P\) is given by
    \begin{equation}
        \frac{\partial U^\star}{\partial P}
        = -\bigl(\mathcal{U}^T \mathcal{Q} \mathcal{U} + \mathcal{R}\bigr)^{-1}
        \mathcal{U}^T \mathcal{Q} \,\mathcal{P}.
    \end{equation}
\end{proposition}
\vspace*{0.2cm}
\begin{proof}
We start by writing the \gls{QP} in \cref{eq:QP_EDS_example} in matrix form and eliminating the equality constraints:
\begin{equation}
    J = (\mathcal{X} x_0 + \mathcal{U} U + \mathcal{P} P)^T \mathcal{Q}  
    (\mathcal{X} x_0 + \mathcal{U} U + \mathcal{P} P) \;+\; U^T \mathcal{R} U.
\end{equation}
Taking the gradient of \( J \) with respect to \( U \) and setting it to zero yields:
\begin{equation}
    \frac{\partial J}{\partial U}
    = 2 \,\mathcal{U}^T \,\mathcal{Q}\, (\mathcal{X} x_0 + \mathcal{U} U + \mathcal{P} P) 
      \;+\; 2 \,\mathcal{R} \, U
    = 0.
\end{equation}
Rearranging, we obtain:
\begin{equation}
    \mathcal{U}^T \,\mathcal{Q}\, (\mathcal{X} x_0 + \mathcal{U} U + \mathcal{P} P) 
      \;+\; \mathcal{R}\, U 
    = 0.
\end{equation}
Solving for \( u \) yields
\begin{equation}
    U^\star 
    = 
    -\bigl(\mathcal{U}^T \mathcal{Q} \mathcal{U} + \mathcal{R}\bigr)^{-1}
    \mathcal{U}^T \mathcal{Q}\,(\mathcal{X} x_0 + \mathcal{P} P),
\end{equation}
where we used that \(\mathcal{U}^T \mathcal{Q} \mathcal{U} + \mathcal{R} \succ 0\) and thus invertible, given that we assumed \(\mathcal{Q} \succeq 0\) and \( R \succ 0 \).
Differentiating \( U^\star \) with respect to \( P \) then gives:
\begin{equation}
    \frac{\partial U^*}{\partial P}
    = 
    -\bigl(\mathcal{U}^T \mathcal{Q} \mathcal{U} + \mathcal{R}\bigr)^{-1}
    \mathcal{U}^T \mathcal{Q} \,\mathcal{P},
\end{equation}
which completes the proof. \qed
\end{proof}

\begin{figure}[tb]
    \centering
    \resizebox{1.0\linewidth}{!}{\input{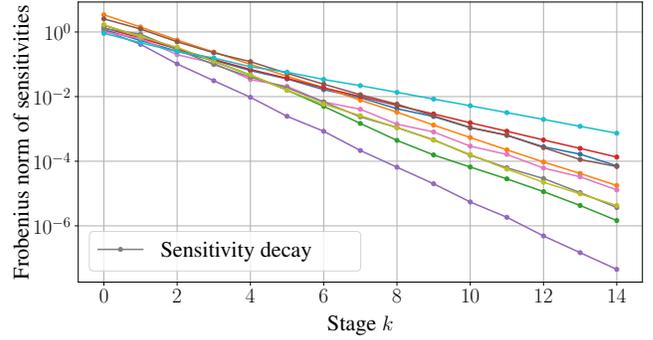}}
    \caption{Frobenius norm of the sensitivity matrix 
    \(\frac{du_0^\star}{dp_k}\) plotted against prediction horizon 
    stages for randomly generated instances of \cref{eq:QP_EDS_example}. Lines of different colors correspond to different problem instances. $S$ is chosen to be the cost-to-go of the corresponding discrete-time \gls{LQR}.}
    \label{fig:sensitivities_motivating_example}
\end{figure}

In \Cref{fig:sensitivities_motivating_example}, we plot the Frobenius norm of the sensitivity matrix \(\left\| \frac{du_0^\star}{dp_k} \right\|_F\) against the stage indices \(k\) for several problem instances of \eqref{eq:QP_EDS_example}, using randomly generated matrices \(Q, R, A,\) and \(B\). We observe that the sensitivities typically decay exponentially with the stage \( k \), indicating that modeling errors in later stages of the horizon have significantly less impact on the initial control input \( u_0 \) compared to errors occurring earlier in the prediction horizon. 

\subsubsection{Previous Work on EDS}
In this subsection, we briefly summarize the results by Shin et~al.~\cite{shin2022exponential}, who establish \gls{EDS} for general graph-structured nonlinear programs, and present them in the context of \gls{MPC}. For simplicity of notation, we neglect the structure of optimal control problems and instead consider the standard form in the following:
\begin{equation}
\label{eq:abstracted_graph_NLP}
\begin{aligned}
    \text{NLP}(P) \coloneqq \; \min_{X} \quad & L(X; P) \\
    \text{s.t.} \quad & C^E(X; P) = 0, \quad & (Y^E) \\
                      & C^I(X; P) \geq 0. \quad & (Y^I)
\end{aligned}
\end{equation}
Here, the capital variables denote the stacked shooting variables and parameters, e.g., \(X := \{x_k, u_k\}_{k=0}^{N}\) and \(P := \{p_k\}_{k=0}^{N}\), where \(x_k\) are the states, \(u_k\) the controls, and \(p_k\) the stage-wise parameters. The function \(L(X; P)\) denotes the overall MPC cost, as introduced in~\cref{eq:baseline_problem}. The parameters \(P\) may appear in the cost, constraints, or dynamics, and can, for instance, represent additive disturbances as in~\Cref{prop:sensitivities_motivating_example} in the previous section. The dual variables $Y^{E}$ and $Y^{I}$ correspond to the equality and inequality constraints, respectively.
We denote the primal-dual solution by \(Z^\star = (X^\star, Y^\star)\) and the corresponding reference parameters by $P^{\star}$. The main \gls{EDS} result is established under the following assumptions:

\begin{assumption}[Twice Continuous Differentiability]
\label{ass:bounded}
The functions \(L: \mathbb{R}^{n_X} \times \mathbb{R}^{n_P} \rightarrow \mathbb{R}\), \(C^E: \mathbb{R}^{n_X} \times \mathbb{R}^{n_P} \rightarrow \mathbb{R}^{n_E}\), and 
\(C^I: \mathbb{R}^{n_X} \times \mathbb{R}^{n_P} \rightarrow \mathbb{R}^{n_I}\) are twice continuously differentiable in a neighborhood of \((X^{\star}, P^{\star})\).
\end{assumption}
\begin{assumption}[LICQ]
\label{ass:LICQ}
At the primal-dual solution \(Z^\star\), the gradients of all equality constraints and active inequality constraints are linearly independent.
\end{assumption}

Intuitively, \gls{LICQ} ensures that a linear approximation to the active constraints captures the local geometry of the feasible region \cite{nocedal1999numerical}. The \gls{SSOSC} imply local convexity in feasible directions, guaranteeing a strict local minimum.

\begin{assumption}[SSOSC]
\label{ass:SSOSC}
Let the Lagrangian be defined as
\begin{equation}
\mathcal{L}(Z; P) := L(X;P) + {Y^E}^\top C^E(X; P) + {Y^I}^\top C^I(X; P).
\end{equation}
Then the \emph{reduced Hessian} at \(Z^\star\), obtained by projecting the Hessian of the Lagrangian onto the null space of the gradients of all equality constraints and active inequality constraints with nonzero dual variables, is positive definite. 
\end{assumption}

Shin et al. establish that, under the stated assumptions, there exist neighborhoods \(\mathbb{P} \subseteq \mathbb{R}^{n_P}\) around \(P^{\star}\) and \(\mathbb{Z} \subseteq \mathbb{R}^{n_Z}\) around \(Z^{\star}\), along with a continuous mapping \(Z^{\dagger}: \mathbb{P} \rightarrow \mathbb{Z}\), such that \(Z^{\dagger}(P)\) remains a well-defined primal-dual solution of \cref{eq:abstracted_graph_NLP} that satisfies \gls{SSOSC} and \gls{LICQ}. The following theorem formalizes \gls{EDS}.

\begin{theorem}[EDS (Theorem 4.5 in \cite{shin2022exponential})]
\label{th:main}
Under assumptions \ref{ass:bounded}, \ref{ass:LICQ} and \ref{ass:SSOSC}, there exists a neighborhood \(\tilde{\mathbb{P}}\subset\mathbb{P}\) of \(P^{\star}\) such that the following holds for any $P, P^{\prime} \in \tilde{\mathbb{P}}$:

\begin{equation}
\left\|z_k^{\dagger}(P) - z_k^{\dagger}(P^{\prime})\right\| 
\leq \sum_{j = 0}^{N} \Upsilon \rho^{\left\lceil\frac{|k - j|}{4} - 1\right\rceil_{+}} 
\left\|p_j - p_j^{\prime}\right\|, \ k \in [N].
\end{equation}
\end{theorem}
\noindent
Here, \(\lceil \cdot \rceil_{+}\) denotes the smallest non-negative integer greater than or equal to the argument. The decay rate \(\rho\) and constant \(\Upsilon\) are scalars that depend on the largest and smallest singular values of submatrices of the Hessian of the Lagrangian and thus, its conditioning. For further details, we refer to \cite{shin2022exponential}. \Cref{th:main} shows that the effect of parametric perturbations on the solution decays exponentially with stage distance.

\subsection{Experiments: further implementation details}

\begin{figure}[t]
  \centering
  \includegraphics[width=1.0\linewidth]{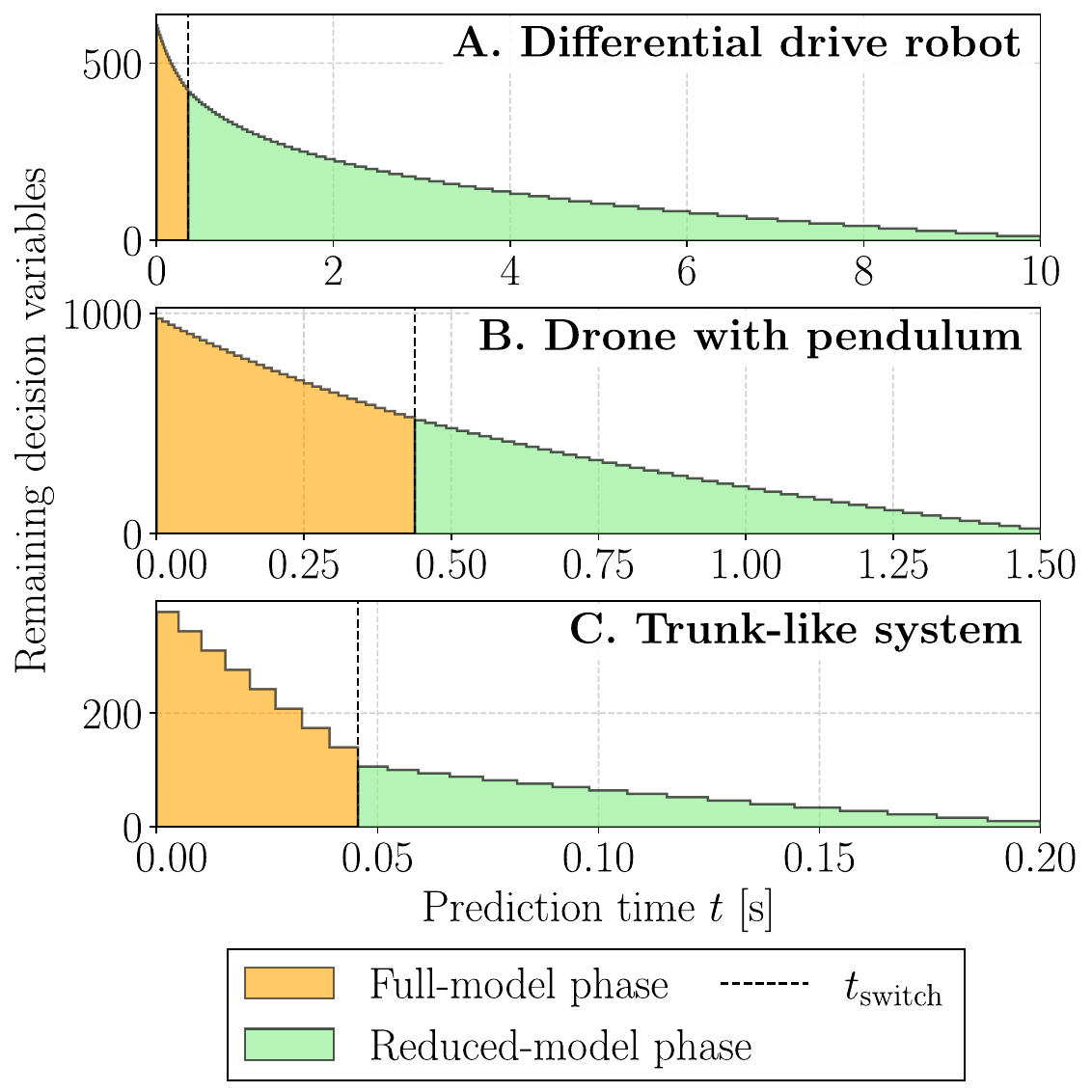}
  \caption{Remaining decision variables versus prediction time. The y-axis shows $\sum_{i=k}^{N-1}(\dim(x_i)+\dim(u_i)) + \dim(x_N)$, i.e., the number of primal state and input variables remaining from stage $k$ onward.}
  \label{fig:remaining_variables_over_time}
\end{figure}

\begin{figure}[t]
  \centering
  \includegraphics[width=1.0\linewidth]{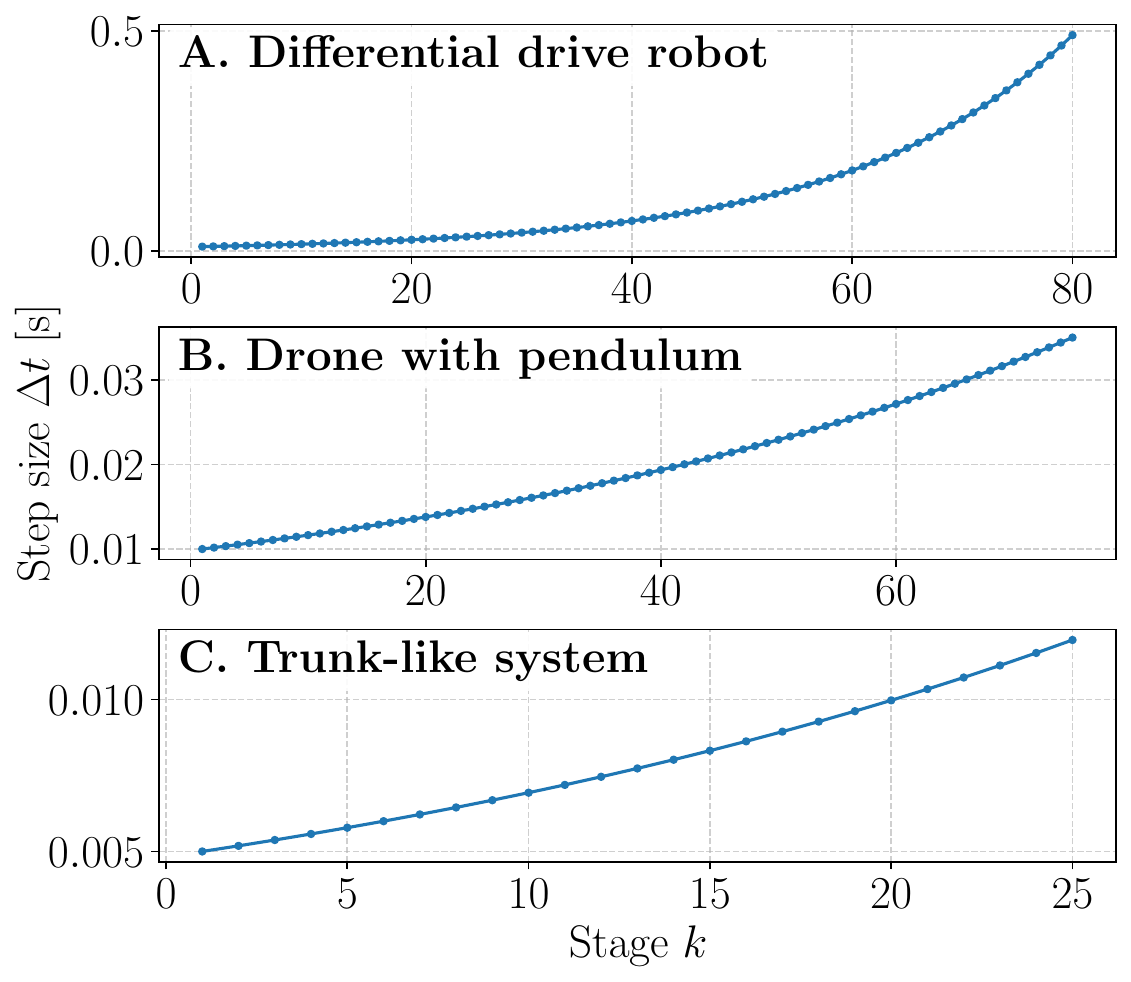}
  \caption{Step size schedule versus stage $k$. The plotted schedules are those used in the experiments.}
  \label{fig:step_sizes_schedule}
\end{figure}

As mentioned, we implement all predictive controllers using the multi-phase interface of the open-source \gls{SQP} framework \texttt{acados} \cite{frey2024multi, verschueren2022acados}, which allows for 
convenient formulation and solution of optimal control problems with multiple phases that may involve different constraints, cost functions, and—crucially for our work—structurally different models. During tuning, we generate and evaluate a separate multi-phase solver for each candidate switching stage $\bar{k}$. In our current implementation, this sweep is not parallelized and, depending on the example, can take several hours. Stage costs are discretized using Euler integration, i.e., each stage cost is weighted by the corresponding time step $\Delta t_k$. 
To visualize how the proposed scheme coarsens the optimization problem along the prediction horizon, \Cref{fig:remaining_variables_over_time} plots the number of remaining decision variables as a function of prediction time. It shows that MTS-MPC preserves a fine resolution near the current time and progressively coarsens the far horizon. The step size schedules used in the experiments are visualized in \Cref{fig:step_sizes_schedule}. 
Example-specific values for $\Delta t_0$, the total number of shooting nodes $N$, and the model-switching stage $\bar{k}$ are listed in  \Cref{tab:mpc_configs_all}. The physical parameters for all examples are listed in \Cref{tab:all_system_parameters}, and performance metrics such as mean stage cost, solve time, time per SQP iteration, and their standard deviations are reported in \Cref{tab:all_results}.

\subsubsection{Oscillation-aware drone control}
For completeness, we report the equations of motion used in the drone-spring-pendulum experiments. The full model is given by
\begingroup\small
\allowdisplaybreaks
\begin{subequations}
\begin{align}
\ddot{p}_y &=
  \frac{
    -\kappa (w_1 + w_2)\sin\varphi
    - k_{S} l_0 \sin\vartheta
    + k_{S} r \sin\vartheta
  }{m_{d}},
  \label{eq:drone_p_y}\\
\ddot{p}_z &=
  \frac{
    -m_{d} g
    + \kappa (w_1 + w_2)\cos\varphi
    + k_{S} l_0 \cos\vartheta
    - k_{S} r \cos\vartheta
  }{m_{d}},
  \label{eq:drone_p_z}\\
\ddot{\varphi} &=
  \frac{
    L_{\text{rot}}\,\kappa\,(-w_1 + w_2)
  }{I_{xx}},
  \label{eq:drone_phi}\\
\ddot{r} &=
  \frac{
    k_{S}(l_0 - r)
  }{m_{l}}
  + r\,\dot{\vartheta}^{2}
  + \frac{
      \kappa (w_1 + w_2)\cos(\varphi - \vartheta)
    }{m_{d}}
  + \frac{
      k_{S}(l_0 - r)
    }{m_{d}},
  \label{eq:drone_r}\\
\ddot{\vartheta} &=
  \frac{
    -2\,m_{d} m_{l}\,\dot{r}\dot{\vartheta}
    + \kappa m_{l}(w_1 + w_2)
      \sin(\varphi - \vartheta)
  }{
    m_{d} m_{l} r
  },
  \label{eq:drone_theta}\\
\dot{w}_1 &= u_1, \\
\dot{w}_2 &= u_2.
\label{eq:drone_w}
\end{align}
\end{subequations}
\endgroup
Here, \(p_y, p_z\) denote the drone position, \(\varphi\) its roll angle, \(r\) the spring length, and \(\vartheta\) the pendulum angle. The quantities \(w_1, w_2\) are rotor speeds, \(m_d\) and \(m_l\) are the drone and load masses, \(I_{xx}\) the body inertia, \(k_S\) and \(l_0\) the spring stiffness and rest length, \(L_{\text{rot}}\) is half the rotor distance, \(\kappa\) the thrust constant, and \(g\) the gravitational acceleration. The drone system is visualized in \Cref{fig:main}-B. The \gls{ODE} of the simplified model, which assumes a rigid pendulum of fixed length~$l_0$, is given by
\begingroup\small
\allowdisplaybreaks
\begin{subequations}
\begin{align}
\ddot{p}_y &=
  \frac{1}{2 m_{d} (m_{d} + m_{l})}
  \Big(
    -2 m_{d}\kappa (w_1 + w_2)\sin\varphi \nonumber\\
&\quad + 2 m_{d} l_0 m_{l} \sin\vartheta\, \dot{\vartheta}^{2} \nonumber\\
&\quad - \kappa m_{l} \big[
      w_1 \sin(\varphi - 2\vartheta) + w_1 \sin\varphi \nonumber\\
&\qquad\qquad\;\; + w_2 \sin(\varphi - 2\vartheta) + w_2 \sin\varphi
    \big]
  \Big),
\\[2pt]
\ddot{p}_z &=
  \frac{1}{2 m_{d} (m_{d} + m_{l})}
  \Big(
    -2 m_{d}^{2} g
    + 2 m_{d}\kappa (w_1 + w_2)\cos\varphi \nonumber\\
&\quad - 2 m_{d} g m_{l}
    - 2 m_{d} l_0 m_{l} \cos\vartheta\, \dot{\vartheta}^{2} \nonumber\\
&\quad - \kappa m_{l} \big[
      w_1 \cos(\varphi - 2\vartheta) - w_1 \cos\varphi \nonumber\\
&\qquad\qquad\;\; + w_2 \cos(\varphi - 2\vartheta) - w_2 \cos\varphi
    \big]
  \Big),
\\[2pt]
\ddot{\varphi} &=
  \frac{L_{\text{rot}}\, \kappa\, (-w_1 + w_2)}{I_{xx}},
\\[2pt]
\ddot{\vartheta} &=
  \frac{\kappa m_{l} (w_1 + w_2)\sin(\varphi - \vartheta)}{m_{d} l_0 m_{l}},
\\[2pt]
\dot{w}_1 &= u_1,
\\
\dot{w}_2 &= u_2.
\end{align}
\end{subequations}
\endgroup

We observed the drone problem to be badly conditioned, occasionally leading to solver failures. To improve robustness, we regularize the Lagrangian by adding a multiple of the identity matrix to its Hessian, as outlined in \cite{nocedal1999numerical}. This regularization technique, often referred to as Levenberg-Marquardt regularization, prevents indefinite Hessians and is supported by \texttt{acados}. However, we observed that this regularization tends to slow down convergence, resulting in increased solve times. As an alternative not considered here, extending the stage cost (\cref{eq:stage_costs}) with a small penalty on additional state components beyond $p_y$ and $p_z$ might also help prevent indefinite Hessians and improve convergence. Finally, we note that we initialize the drone in hover, with the spring pendulum at rest length $l_0$. 

\begin{table*}[h]
  \centering
  \small
  \setlength{\tabcolsep}{6pt}
  \renewcommand{\arraystretch}{1.2}
  \begin{tabular}{llll}
    \toprule
    \textbf{System} & \textbf{Model / Phase} & \textbf{Matrix} & \textbf{Value} \\
    \midrule

    \multirow{3}{*}{\textbf{Differential Drive}}
      & Full (with actuators)   & $Q$       & $\mathrm{diag}(10^{4},\, 10^{4},\, 10^{-4},\, 5\!\cdot\!10^{3},\, 10^{-3},\, 0.5,\, 0.5)$ \\
      & Simplified              & $\hat{Q}$ & $\mathrm{diag}(10^{4},\, 10^{4},\, 10^{-4},\, 5\!\cdot\!10^{3},\, 10^{-3})$ \\
      & Simplified              & $\hat{R}$ & $\mathrm{diag}(0.5,\, 0.5)$ \\

    \midrule
    \multirow{3}{*}{\textbf{Drone}}
      & Full \& Approx.         & $Q$            & $\mathrm{diag}(1.0,\, 1.0)$ \\
      & Full \& Approx.         & $Q_{\text{load}}$ & $\mathrm{diag}(0.1,\, 0.1)$ \\
      & Full \& Approx.         & $R$            & $\mathrm{diag}(10^{-3},\, 10^{-3})$ \\

    \midrule
    \multirow{3}{*}{\textbf{Trunk-like}}
      & Full model              & $Q$ & $\mathrm{diag}(2,\, 2,\, 0.025,\, 0.025)$ \\
      & Full model              & $R$ & $0.5 \cdot \mathrm{diag}(0.015,\, 0.015)$ \\
      & Point-mass (approx.)    & $R$ & $\mathrm{diag}(5\!\cdot\!10^{-6},\, 5\!\cdot\!10^{-6})$ \\

    \bottomrule
  \end{tabular}
  \caption{Weighting matrices used in the stage/terminal costs across all systems.}
  \label{tab:q_r_values}
\end{table*}

\begin{table*}[h]
  \centering
  \small
  \setlength{\tabcolsep}{6pt}
  \renewcommand{\arraystretch}{1.2}
  \begin{tabular}{llccc}
    \toprule
    \textbf{System} & \textbf{Approach} &
    \textbf{Mean Cost} &
    \textbf{Solve Time [s]} &
    \textbf{Time / SQP-Iter [s]} \\
    \midrule

    \multirow{6}{*}{\textbf{Diff. Drive}}
      & 0) Baseline                     & $4705.448\ (\pm\ 12928.892)$ & $0.237\ (\pm\ 0.357)$ & $0.017\ (\pm\ 0.016)$ \\
      & 1) Shorter Horizon              & $4955.323\ (\pm\ 12642.376)$ & $0.029\ (\pm\ 0.040)$ & $0.004\ (\pm\ 0.004)$ \\
      & 2) Larger Step Size             & $4736.789\ (\pm\ 12929.419)$ & $0.068\ (\pm\ 0.105)$ & $0.004\ (\pm\ 0.004)$ \\
      & 4) Increasing Step Sizes        & $4705.674\ (\pm\ 12934.705)$ & $0.020\ (\pm\ 0.031)$ & $0.001\ (\pm\ 0.001)$ \\
      & 5) Model Switching              & $4705.710\ (\pm\ 12925.564)$ & $0.132\ (\pm\ 0.203)$ & $0.009\ (\pm\ 0.009)$ \\
      & 6) Model Switching + Inc. Steps & $4707.832\ (\pm\ 12923.114)$ & $0.015\ (\pm\ 0.023)$ & $0.001\ (\pm\ 0.001)$ \\

    \midrule
    \multirow{7}{*}{\textbf{Drone}}
      & 0) Baseline                     & $25.070\ (\pm\ 40.842)$ & $0.627\ (\pm\ 1.217)$ & $0.016\ (\pm\ 0.027)$ \\
      & 1) Shorter Horizon              & $26.661\ (\pm\ 43.050)$ & $0.424\ (\pm\ 1.250)$ & $0.010\ (\pm\ 0.017)$ \\
      & 2) Larger Step Size             & $26.043\ (\pm\ 41.446)$ & $0.279\ (\pm\ 0.374)$ & $0.005\ (\pm\ 0.006)$ \\
      & 3) Approx. Model      & $29.428\ (\pm\ 40.149)$ & $0.324\ (\pm\ 0.524)$ & $0.009\ (\pm\ 0.009)$ \\
      & 4) Increasing Step Sizes        & $25.132\ (\pm\ 41.088)$ & $0.204\ (\pm\ 0.262)$ & $0.005\ (\pm\ 0.005)$ \\
      & 5) Model Switching              & $25.143\ (\pm\ 41.124)$ & $0.346\ (\pm\ 0.459)$ & $0.009\ (\pm\ 0.009)$ \\
      & 6) Model Switching + Inc. Steps & $25.142\ (\pm\ 41.052)$ & $0.189\ (\pm\ 0.301)$ & $0.004\ (\pm\ 0.005)$ \\

    \midrule
    \multirow{6}{*}{\textbf{Trunk-like}}
      & 0) Baseline                     & $0.00972\ (\pm\ 0.01283)$ & $0.089\ (\pm\ 0.012)$ & $0.031\ (\pm\ 0.002)$ \\
      & 1) Shorter Horizon              & $0.00993\ (\pm\ 0.01356)$ & $0.052\ (\pm\ 0.008)$ & $0.019\ (\pm\ 0.001)$ \\
      & 2) Larger Step Size             & $0.00985\ (\pm\ 0.01271)$ & $0.048\ (\pm\ 0.008)$ & $0.014\ (\pm\ 0.001)$ \\
      & 4) Increasing Step Sizes        & $0.00975\ (\pm\ 0.01270)$ & $0.054\ (\pm\ 0.010)$ & $0.018\ (\pm\ 0.001)$ \\
      & 5) Model Switching              & $0.00973\ (\pm\ 0.01301)$ & $0.021\ (\pm\ 0.003)$ & $0.008\ (\pm\ 0.001)$ \\
      & 6) Model Switching + Inc. Steps & $0.00973\ (\pm\ 0.01299)$ & $0.017\ (\pm\ 0.002)$ & $0.006\ (\pm\ 0.0005)$ \\

    \bottomrule
  \end{tabular}
  \caption{Mean values (\(\pm\) standard deviation) of stage cost, total solve time, and solve time per SQP iteration for each example system and MPC variant.}
  \label{tab:all_results}
\end{table*}

\begin{table*}[h]
  \centering
  \small
  \setlength{\tabcolsep}{6pt}
  \renewcommand{\arraystretch}{1.2}
  \begin{tabular}{llccccc}
    \toprule
    \textbf{System} & \textbf{Approach} &
    $\boldsymbol{\Delta t_0}$ [s] & $\boldsymbol{N}$ &
    $\boldsymbol{\bar{k}}$ & $\boldsymbol{T}$ [s] & $\boldsymbol{f}$ [Hz] \\
    \midrule

    \multirow{6}{*}{\textbf{Differential Drive}}
      & 0) Baseline                     & 0.01 & 1000   & -- & 10.0  & 100 \\
      & 1) Shorter Horizon              & 0.01 & 250    & -- & 2.5   & 100 \\
      & 2) Larger Step Size             & 0.04 & 250    & -- & 10.0  & 25  \\
      & 4) Increasing Step Sizes        & 0.01 & 80     & -- & 10.0  & 100 \\
      & 5) Model Switching              & 0.01 & 1000   & 36 & 10.0  & 100 \\
      & 6) Model Switching + Inc. Steps & 0.01 & 80     & 21 & 10.0  & 100 \\

    \midrule
    \multirow{7}{*}{\textbf{Drone}}
      & 0) Baseline                     & 0.01 & 150  & -- & 1.50 & 100 \\
      & 1) Shorter Horizon              & 0.01 & 100  & -- & 1.00 & 100 \\
      & 2) Larger Step Size             & 0.02 & 75   & -- & 1.50 & 50  \\
      & 3) Approx. Model               & 0.01 & 150  & -- & 1.50 & 100 \\
      & 4) Increasing Step Sizes        & 0.01 & 75   & -- & 1.50 & 100 \\
      & 5) Model Switching              & 0.01 & 150  & 45 & 1.50 & 100 \\
      & 6) Model Switching + Inc. Steps & 0.01 & 75   & 33 & 1.50 & 100 \\

    \midrule
    \multirow{6}{*}{\textbf{Trunk-like}}
      & 0) Baseline                     & 0.005 & 40      & -- & 0.200 & 200 \\
      & 1) Shorter Horizon              & 0.005 & 25      & -- & 0.125 & 200 \\
      & 2) Larger Step Size             & 0.010 & 20      & -- & 0.200 & 100 \\
      & 4) Increasing Step Sizes        & 0.005 & 25      & -- & 0.200 & 200 \\
      & 5) Model Switching              & 0.005 & 40      & 10 & 0.200 & 200 \\
      & 6) Model Switching + Inc. Steps & 0.005 & 25      & 8 & 0.200 & 200 \\

    \bottomrule
  \end{tabular}
  \caption{MPC configuration parameters. Columns list the initial step size $\Delta t_0$, prediction horizon $N$, switching index $\bar{k}$ (if used), time look-ahead $T$, and control frequency $f = 1/\Delta t_0$.}
  \label{tab:mpc_configs_all}
\end{table*}

\begin{table*}[h]
  \centering
  \small
  \setlength{\tabcolsep}{6pt}
  \renewcommand{\arraystretch}{1.2}
  \begin{tabular}{lll}
    \toprule
    \textbf{System} & \textbf{Parameter} & \textbf{Value} \\
    \midrule

    \multirow{10}{*}{\textbf{Differential Drive}}
      & Total mass $m_{\mathrm{tot}}$                    & $220\,\mathrm{kg}$ \\
      & Mass without wheels $m_{c}$                      & $200\,\mathrm{kg}$ \\
      & Wheel radius $R_{w}$                             & $0.16\,\mathrm{m}$ \\
      & Half track width $L_{tw}$                        & $0.32\,\mathrm{m}$ \\
      & Rear-axle distance $d$                           & $0.01\,\mathrm{m}$ \\
      & Body inertia $I_{zz}$                            & $9.6\,\mathrm{kg\,m^{2}}$ \\
      & Wheel\,+\,motor inertia $I_{w}$                  & $0.1\,\mathrm{kg\,m^{2}}$ \\
      & Motor constants $K_{1},K_{2}$                    & $1.0$ \\
      & Actuator inductance $L_{\mathrm{act}}$           & $10^{-4}\,\mathrm{H}$ \\
      & Actuator resistance $R_{\mathrm{act}}$           & $0.05\,\Omega$ \\

    \midrule
    \multirow{8}{*}{\textbf{Drone}}
      & Drone mass $m_{d}$                               & $2.0\,\mathrm{kg}$ \\
      & Load mass $m_{l}$                                & $0.3\,\mathrm{kg}$ \\
      & Body inertia $I_{xx}$                            & $0.05\,\mathrm{kg\,m^{2}}$ \\
      & Gravity $g$                                      & $9.81\,\mathrm{m/s^{2}}$ \\
      & Spring stiffness $k_{S}$                         & $2000\,\mathrm{N/m}$ \\
      & Spring rest length $l_{0}$                       & $0.3\,\mathrm{m}$ \\
      & Rotor thrust constant $\kappa$                   & $1.0$ \\
      & Half rotor distance $L_{\mathrm{rot}}$           & $0.2\,\mathrm{m}$ \\

    \midrule
    \multirow{7}{*}{\textbf{Trunk-like}}
      & Gravity $g$                                      & $9.81\,\mathrm{m/s^{2}}$ \\
      & Link mass $m_{l}$                                & $2.98\times10^{-3}\,\mathrm{kg}$ \\
      & Link half length $l$                             & $0.015625\,\mathrm{m}$ \\
      & Link inertia $I_{zz}$                            & $3.79\times10^{-7}\,\mathrm{kg\,m^{2}}$ \\
      & Joint stiffness $c$                              & $3\,\mathrm{N\,m/rad}$ \\
      & Joint damping $d_{l}$                            & $0.15\,\mathrm{N\,m\,s/rad}$ \\
      & Gear constant                                    & $2$ \\

    \bottomrule
  \end{tabular}
  \caption{Physical parameters.}
  \label{tab:all_system_parameters}
\end{table*}

\end{document}